\documentclass[twocolumn,showpacs]{revtex4-1}

\usepackage{graphicx}
\usepackage{dcolumn}
\usepackage{bm}
\usepackage{amsmath}  
\usepackage[abs]{overpic}

\usepackage{aas_macros}

\begin{document}

\bibliographystyle{plainnat}





\title{Kinetic turbulence in space plasmas observed in the near-Earth and near-Sun solar wind.}

\author{Olga Alexandrova\textsuperscript{1}, Vamsee Jagarlamudi\textsuperscript{2,1}, Claudia Rossi\textsuperscript{1}, \\
Milan Maksimovic\textsuperscript{1}, Petr Hellinger\textsuperscript{3,4}, Yuri Shprits\textsuperscript{5}, \& Andre Mangeney\textsuperscript{1}}
\affiliation{\textsuperscript{1}LESIA, Observatoire de Paris, Universit\'e PSL, CNRS, Sorbonne Universit\'e, Universit\'e de Paris, 5 place Jules Janssen, 92195 Meudon, France.}
\affiliation{\textsuperscript{2}LPC2E, CNRS, University of Orl\'eans, 3 Avenue de la Recherche Scientifique, 45071 Orleans Cedex 2, France.}
\affiliation{\textsuperscript{3}Astronomical Institute, CAS, Bocni II/1401, Prague CZ-14100, Czech Republic}
\affiliation{\textsuperscript{4}Institute of Atmospheric Physics, CAS, Bocni II/1401, Prague CZ-14100, Czech Republic}
\affiliation{\textsuperscript{5}GFZ German Research Centre for Geosciences, University of Potsdam, Germany.}

\date{March 2, 2020}

\begin{abstract}

Turbulence develops in any stressed flow when the scales of the forcing are much larger than those of the dissipation. In neutral fluids, it consists of chaotic motions in physical space but with a universal energy spectrum in Fourier space. Intermittency (non-Gaussian statistics of fluctuations) is another general property and it is related to the presence of  coherent structures. Space plasmas are turbulent as well. Here, we focus on the kinetic plasma scales, which are not yet well understood. We address the following  fundamental questions: (1)~Do the turbulent fluctuations at kinetic scales form a universal spectrum? and (2)~What is the nature of the fluctuations? Using measurements in the solar wind we show that the magnetic spectra of  kinetic turbulence at 0.3, 0.6 and 0.9~AU from the Sun have the same shape as the ones close to the Earth orbit at 1~AU, indicating universality of the phenomenon. The fluctuations, which form this spectrum, are typically non-linearly interacting eddies that tend to generate magnetic filaments.

\end{abstract}

\maketitle

\section{Current state of the field and long-standing questions}

Turbulence is one of the unsolved problems in physics: to date, there is no satisfactory theory, based on first principles, that describes turbulence in a sufficiently general frame. Therefore, one has to rely on phenomenologies. For example, the observed general spectrum of the  inertial range of the incompressible neutral fluid turbulence $\sim k^{-5/3}$  (with wavenumber $k$) is well described by  
Kolomogorov's phenomenology \cite{k41}. Intermittency   (or non-Gaussianity of fluctuations and their dependency on scale) is beyond this description, and in neutral fluids it  is due to coherent structures  like filaments of vorticity. Their cross section is of the order of the dissipation scale $\ell_d$.  The dissipation range is described by another general law $\sim k^{-3}\exp{(-Ck\ell_d)}$, with $C$ being a constant close to 7  \cite{chen93}. Such a spectrum with an exponential correction indicates a lack of self-similarity in the dissipation range where turbulent energy is transfered into heat.

Astrophysical plasmas differ from usual neutral fluids. Natural plasmas 
(i)  are almost collisionless so that the viscosity and Kolomogorov's dissipation scale $\ell_d$ are ill-defined; 
(ii) are almost completely ionized so that  the presence of a  magnetic field will introduce an anisotropy and allow waves to propagate, even in the incompressible limit  (Alfv\'en waves);
(iii) are characterized by a  number of characteristic plasma (or kinetic) scales; 
(iv) are dispersive:  beyond Alfv\'en waves,  one may also expect  fast and/or slow magnetosonic waves at magnetohydrodynamic (MHD) scales, and, at kinetic scales, kinetic Alfv\'en, whistler or slow/ion-acoustic waves, etc...

Considering all this complexity, one may wonder if there is a certain degree of generality in space plasma turbulence. If this is the case, are there similarities with incompressible neutral fluid turbulence? To answer these questions, the solar wind plasma, which is accessible to in-situ space exploration, has proven to be a  very useful laboratory. However,  it is  inhomogeneous, with a dense slow wind blowing at low heliographic latitudes and a fast and more tenuous wind at high latitudes; it is also in  spherical expansion, so that some extra complexity is added: in particular the particle distribution functions tend to develop  strong anisotropies and become unstable, generating fluctuations at small, kinetic scales. 
 
Since the first early in-situ measurements, \cite[e.g.,][]{Coleman1968},our knowledge of the large-scale turbulence in the solar wind has greatly improved, \cite[e.g.,][]{bruno13,kiyani15}. There is an extended inertial range of scales, where  incompressible MHD phenomenologies \cite{goldreich95,boldyrev05,chandran15}, similar in spirit to  the Kolomogorov's phenomenogy, may be invoked to understand the formation  of a Kolmogorov-like spectrum of magnetic fluctuations $\sim k^{-5/3}$~\footnote{Satellite measurements are time series. Thus, in Fourier space one gets frequency spectra. As far as any characteristic plasma velocity, except whistlers waves phase speed, is less than the solar wind speed $V$, one can invoke the Taylor’s hypothesis and convert a spacecraft-frame frequency $f$ to a flow-parallel wavenumber $k$ in the plasma frame $k=2\pi f/V$.}.
At the short wavelength end of the inertial domain, i.e., at scales of the order of the ion inertial scale $\lambda_p=c/\omega_{pp}$ (where $c$ is the speed of light  and $\omega_{pp}$ is the proton plasma frequency) the spectrum steepens. At these scales ($\sim 100$~km at 1~AU), the MHD approximation is no longer valid; the ``heavy''  ion (basically proton in the solar wind) fluid and the ``light'' electron fluid behave separately,  \cite[e.g.,][]{Matthaeus2008PRL,Hellinger2018ApJL,Papini2019ApJ}. 
At even smaller scales, at the vicinity of  the electron scales ($\sim 1$~km at 1~AU), the fluid description does not hold any more  and electrons should be considered as particles. The present paper is concerned  with this short wavelength range, i.e., between ion scales and a fraction of electron scales.

Recently, thanks to a very sensitive Search Coil Magnetometer on the ESA/Cluster mission \cite{escoubet97,cornilleau-wehrlin97},   the small scale tail of the electromagnetic cascade could be explored  down to  electron scales, i.e., the electron inertial length $\lambda_e=c/\omega_{ep}$ (where $\omega_{ep}$ is the electron plasma frequency),  the electron Larmor radius $\rho_e = V_{e\perp}/\omega_{ce}$ (where $V_{e\perp} = \sqrt{2kT_{e\perp}/m_e}$ is the electron thermal speed in the plane perpendicular to the mean magnetic field ${\bf B}_0$, $T_{e\perp}$ is the perpendicular to  ${\bf B}_0$ electron temperature, $m_e$ is the mass of electron and $\omega_{ce}=2\pi f_{ce}$ is the electron cyclotron frequency) and below ($\sim 0.2 - 1$~km)  \cite{alexandrova09,alexandrova12,alexandrova13, sahraoui10,sahraoui13}. 

The Cluster mission operates at 1~AU, and provides observations which seem confusing at first glance. At electron scales, the spectral shape of the magnetic fluctuations vary from event to event  suggesting  that the spectrum is not universal at kinetic scales \cite{sahraoui10,sahraoui13}. However, it may be shown  \cite{roberts17} that most of these spectral variations are due to the presence or absence of  whistler waves  with frequencies of a fraction of $f_{ce}$ and wave vectors ${\bf k}$ quasi-parallel to ${\bf B_0}$ \cite{lacombe14}. These waves may result from the development of some instabilities associated either to an increase of  the electron temperature anisotropy or an increase of the electron heat flux, in some regions of the solar wind  \cite{Stverak2008JGR}. 
 
Are whistler waves part of the background turbulence at kinetic scales? 
Turbulent fluctuations at these scales have low frequencies in the plasma frame ($f\simeq 0$) and wave-vectors mostly perpendicular to the mean field ${\bf k} \perp {\bf B_0}$~\cite{lacombe17}. This background turbulence is convected by the solar wind  (with the speed {\bf V}) across the spacecraft and appears in the satellite frame at frequencies $f=k_{\perp} V/2\pi$. It happens that these frequencies are below but close to $f_{ce}$, exactly in the range where whistler waves (with ${\bf k}{\|}{\bf B_0}$ and $f\le f_{ce}$) may appear locally. Therefore, the superposition of turbulence and whistlers at the same frequencies is incidental. If we could do measurements directly in the plasma frame, these two phenomena  would be completely separated in ${\bf k}$ and $f$. A possible interaction between turbulence and whistlers is out of the scope of the present paper. We focus here on the background turbulence at kinetic scales only.

A statistical study of solar wind streams at 1~AU under different plasma conditions has shown that, in absence of  whistler waves, the turbulent spectrum seems to follow a general shape $\sim f^{-8/3}\exp{(-f/f_d)}$ \cite{alexandrova12}. The characteristic frequency $f_d$ is strongly correlated with the Doppler-shifted electron Larmor radius  $f_{\rho_e}=V/(2\pi \rho_e)$ \cite{alexandrova09,alexandrova12},  and, by analogy with the neutral fluid case, it is referred  here to as the ``dissipation frequency'', associated with  a local ``dissipation scale'' $\ell_d = V/(2\pi f_d)$. 

Two basic questions arise.  First question: how general  is the kinetic spectrum, observed at 1~AU \cite{alexandrova12}? 
Intuitively, one may think that turbulent fluctuations at such small scales are well mixed, showing the same spectrum at any radial distance from the Sun, except for the temporary excitation of whistler waves. We address this question in section II, where we analyse the data from Helios, the first mission which explored the inner Heliosphere up to the orbit of Mercury ($0.3$~AU). Helios provided magnetic spectra  up to the electron scales  at radial distances $0.3-1$~AU \cite{denskat83}.     
These measurements  have shown that   the mean magnetic spectrum at 0.3~AU follows a power-law $\sim f^{-3}$ between ion and electron scales,  but  a global and systematic analysis of  the spectral behaviour of the magnetic field fluctuations over the whole accessible scale range is still lacking, except at 1~AU. We present  an analysis  of  more than 240 000 available magnetic spectra of kinetic turbulence in the inner Heliosphere obtained by Helios-1 in three distance range: around  $0.3$, $0.6$ and $0.9$~AU. We show that these spectra follow the general shape previously observed at 1~AU \cite{alexandrova12}. 

 Second question: what is the nature of fluctuations at kinetic scales? 
 \textit{Weak} (or \textit{wave}) turbulence implies a mixture of small amplitude \textit{weakly interacting random phase waves}, which follow a particular dispersion relation. 
In the case of weak  turbulence, the dissipation is homogeneous in space and is due to collisionless dissipation of waves, such as Landau damping, \cite[e.g.,][]{Laveder2013AnGeo}. Since, the amplitudes of fluctuations at kinetic scales are small in the solar wind,  the wave turbulence scenario is frequently assumed; but  an unambiguous identification of the  type of waves (and the corresponding dispersion relation) in the kinetic range remains elusive, despite  a vast amount of work encompassing theory, simulations and observations aimed at resolving this `wave-debate’, \cite[e.g.,][]{denskat83,galtier06b,sahraoui10,camporeale11,cerri16, roberts13,narita16}. 
On the other hand, \textit{strong turbulence} implies non-linearly interacting eddies leading to the generation of coherent structures (energetic events with coupled phases across a wide range of scales). Then, the dissipation is non-homogeneous in space and concentrates within or near coherent structures. 

 We address the second question in section III. Using Cluster/STAFF time-domain measurements at 1~AU, we show that turbulence at kinetic scales is dominated by coherent structures in the form of magnetic vortices (or current filaments), similar to what is observed in the inertial range of the solar wind turbulence and in usual fluid turbulence.


\section{Spectrum of the kinetic plasma turbulence} 


\begin{figure*}
\begin{overpic}[scale=0.55]{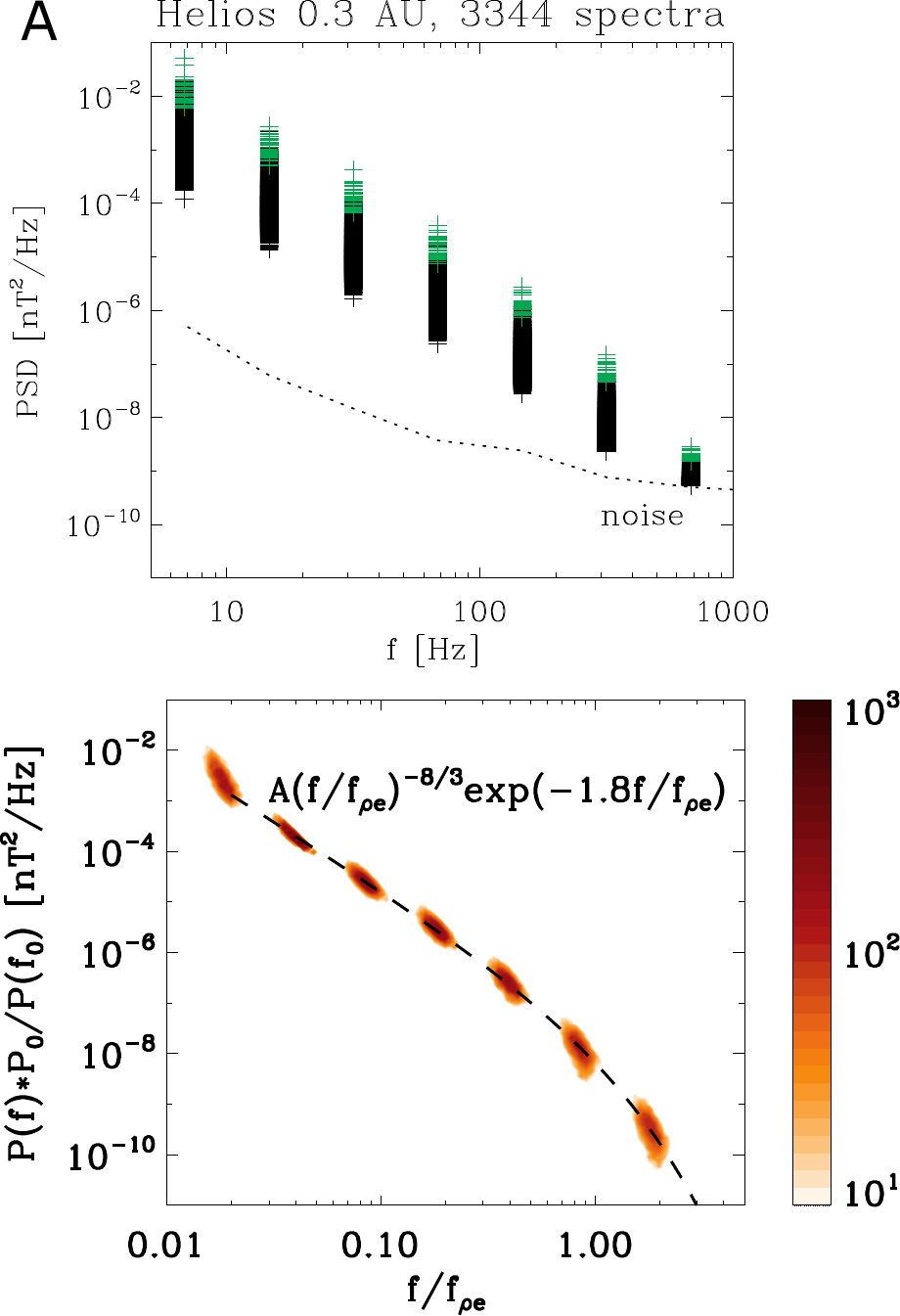}
\end{overpic}
\hspace{1cm}
\begin{overpic}[scale=0.55]{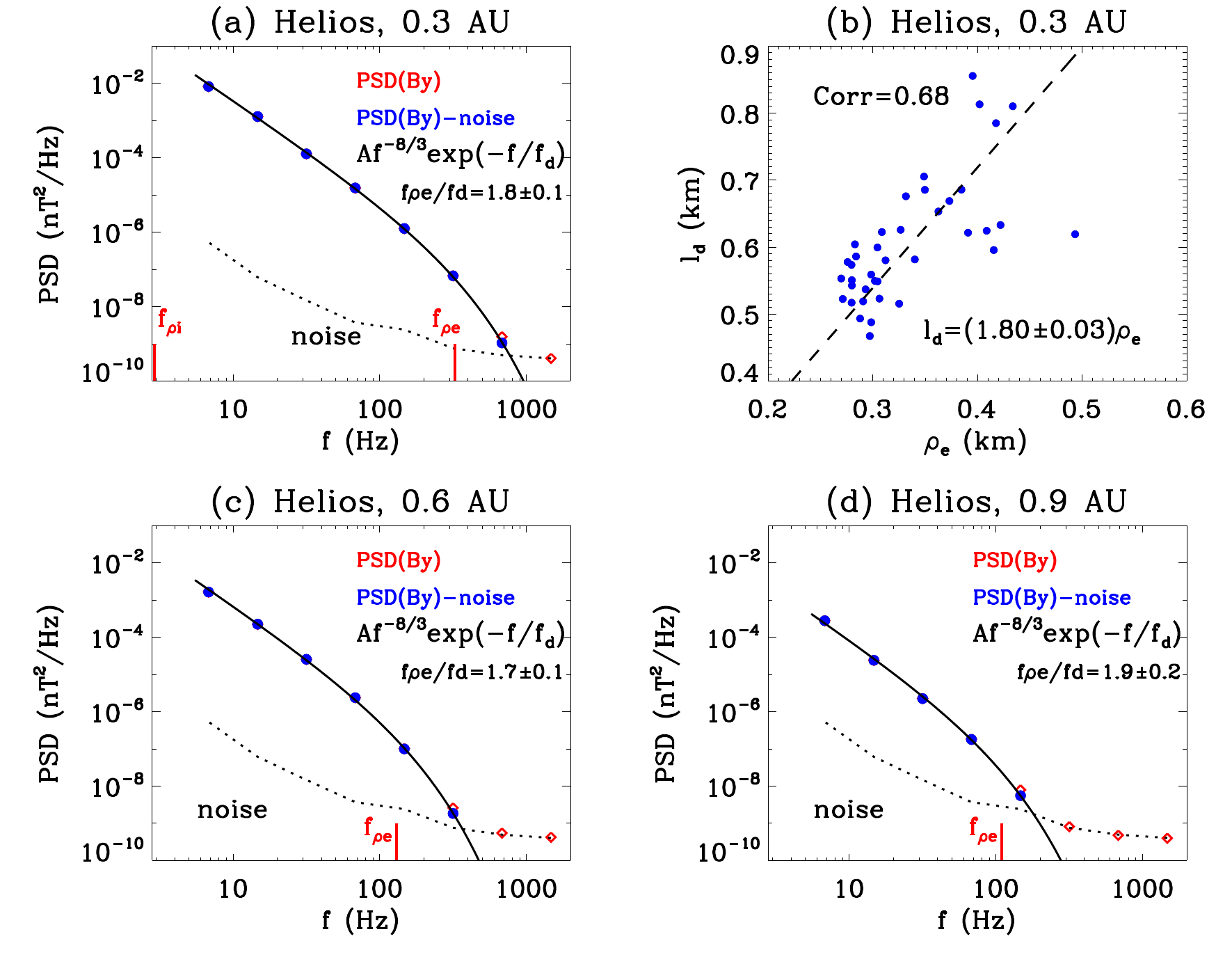}
\put(0,210){\textsf{B}}
\end{overpic}
\caption{ A(top) 3344 individual spectra of $B_y$ at 0.3 AU in the fast wind as measured by Helios-1/SCM; the 39 most intense spectra  are marked by green crosses; SCM noise is indicated by a dotted line. 
A(bottom): The same 3344 spectra, but  cleaned from the noise, normalised by $f_{\rho e}$ and collapsed in amplitude  at $\sim0.05f/f_{\rho e}$; the result is shown as a 2D histogram with the number of the data points proportional to the darkness of the orange colour. The dashed line indicates the function $A(f/f_{\rho e})^{-8/3}\exp(-1.8f/f_{\rho e})$ which passes nicely through the data. B(a): One example of the most intense spectrum at 0.3~AU; the raw-spectrum is shown by red diamonds, the cleaned spectrum, after the subtraction of the noise -- by blue dots; vertical red lines give the Doppler shifted $\rho_p$ and $\rho_e$ appearing at $f_{\rho p} = 2.9$~Hz and $f_{\rho e}= 325$~Hz respectively;  black solid line gives the fit with the model function eq.(\ref{eq:model-2param}).   
B(b) Results of the fitting procedure of the most intense spectra with eq.(\ref{eq:model-2param}): dissipation scale $\ell_d = V/2\pi f_d$ as a function of the electron Larmor radius $\rho_e$; the linear dependence $\ell_d = 1.8 \rho_e$ is indicated by the dashed-line, with the correlation coefficient being $Corr=0.68$. 
B(c) and B(d) The same as B(a) but at 0.6 and 0.9~AU respectively, at both distances, $f_{\rho p}\simeq 1$~Hz.  
At all these radial distances from the Sun, the turbulent spectrum follows the same shape $\sim f^{-8/3}\exp{(-Cf/f_{\rho e})}$, indicating independence of the solar wind expansion. }
\label{fig1}
\end{figure*}

To check  whether the magnetic spectrum at kinetic scales displays a general shape independent of the distance $R$ from the Sun, we have analysed  246543 individual magnetic spectra of Helios-1/SCM for radial distances $R \in [0.3,0.9]$~AU. These spectra satisfy signal-to-noise ratios (SNR) of 2 up to 100~Hz. Among them,  about $ 2\%$  of the spectra show signatures of whistler waves as in \cite{lacombe14}. These waves are mostly present in the slow wind and their occurrence as a function of $R$ and solar wind speed is described in \cite{Jagarlamudi2020ApJ}. The spectra with signatures of whistlers were eliminated from the present study. 

At 0.3 AU, in the fast wind we do not observe any whistler waves. 
At this radial distance, there are 3344 spectra with SNR$>3$ up to 316~Hz  and for which it was possible to determine plasma parameters, see Figure~\ref{fig1}A(top).
All the spectra have similar shape and  
they are similar to what is observed at 1~AU \cite{alexandrova12}. 
Among this dataset, we select the 39 most intense spectra  (see green crosses in Figure~\ref{fig1}A(top)) with SNR$>3$ up to $681$~Hz and with  simultaneous measurements of $\mathbf{B_0}$. One of these spectra is shown in Figure~\ref{fig1}B(a). 
We performed a least square fit  of these best-resolved spectra  against a model with three free parameters:
 \begin{equation}\label{eq:model}
P_{A,\alpha,f_d}\colon f\mapsto \; Af^{-\alpha}\exp{(-f/f_d)}.
\end{equation}
 We found  that the spectral index $\alpha$  is distributed around $8/3$, but   coupled with $f_d$ (this reflects the fact that changing $\alpha$, the fitting procedure changes $f_d$). 
Moreover, the Doppler-shifted dissipation scale $\ell_d = V/2\pi f_d$ correlates with the electron Larmor radius $\rho_e$ with a correlation coefficient of $0.6$.  
Then, we fix $\alpha = 8/3$, and we redo the fitting of the observed spectra with the  
 two-parameter model function: 
 \begin{equation}\label{eq:model-2param}
P_{A,f_d}\colon f\mapsto \; Af^{-8/3}\exp{(-f/f_d)}.
\end{equation}
The correlation between $\ell_d$ and $\rho_e$  increases  slightly ($Corr \sim 0.7$)  and the relation  $\ell_d \sim 1.8 \rho_e$  is observed, see Figure~\ref{fig1}B(b). There is no correlation with the electron inertial length $\lambda_e$ ($Corr = 0.02$, not shown). 
Finally, we compare the 3344 spectra with the one-parameter model
\begin{equation}\label{eq:model-1param}
P_{A}\colon f\mapsto \; Af^{-8/3}\exp{(-1.8 f/f_{\rho e})}.
\end{equation}
The observed spectra are well described by this model, see  Figure~\ref{fig1}A(bottom).

 
\begin{figure*}
\includegraphics[width=10cm]{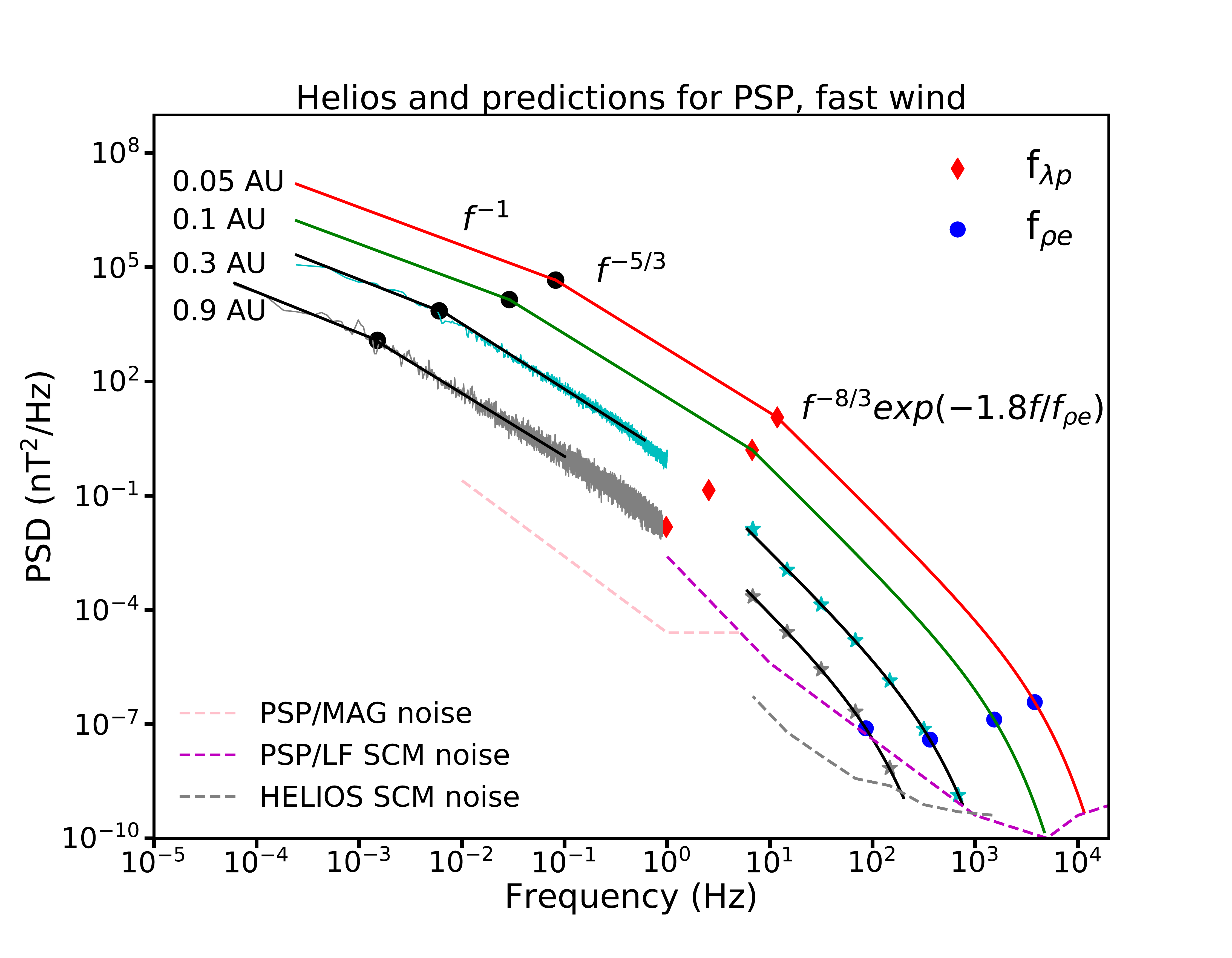} 
\caption{
The complete turbulent spectrum from energy injection scales  up to the sub-electron scales at 4 radial distances from the Sun: Helios measurements at 0.9 (grey) and 0.3~AU (light blue); the predictions for the Parker Solar Probe measurements at 0.1 (green) and 0.05~AU (red). 
The injection scales (which correspond to $\sim f^{-1}$ spectrum) and the MHD inertial range 
($\sim f^{-5/3}$) at 0.9 and 0.3~AU are covered by the Helios/MAG instrument. The Helios/SCM instrument covers the kinetic scales (stars), studied in the present paper. Dashed lines indicate noise levels of the different magnetic sensors on Helios and PSP. 
The Doppler shifted ion inertial length $\lambda_p$ (red diamonds) marks the transition from the inertial to the kinetic range;  the electron Larmor radius $\rho_e$ (blue dots) marks the dissipation range. 
}
\label{fig:psp}
\end{figure*}

Further from the Sun, as expected, the intensity of the spectra decreases with $R$, see Figure~\ref{fig1}B(c) and (d). Regarding the shape of the spectra, the spectral index does not change ($\alpha=8/3$) and the dissipation frequency $f_d$ decreases (from $\sim180$~Hz at 0.3~AU to $\sim 55$~Hz at $0.9$~AU) following $f_{\rho e}$. 
These observations,  namely the spectral shape and the correlation between $f_d$
 and $f_{\rho e}$, agree with our results at 1~AU \cite{alexandrova12}. 
 Thus, the turbulent spectrum at plasma kinetic scales follows the same shape at different radial distances from the Sun  and it is not sensitive to the solar wind expansion, at least between $0.3$ and 1~AU.


 There are only few theoretical and numerical studies showing turbulent spectrum with an exponential roll-off at electron scales \cite{howes11pop, Schreiner2017, Parashar2018ApJL}. The gyrokinetic model of \citet{howes11pop} assumes a  critically balanced kinetic turbulence and the dissipation on the electrons via Landau damping. A similar approach is chosen in the analytical model of \citet{Schreiner2017}. \citet{Parashar2018ApJL} simulate plasma with fully kinetic PIC code. There, the spectral curvature at electron scales seems to depend on $\beta_e$, which is equivalent to the $\rho_e$ dependency observed here.

 
Our results allow us to suggest a form of the turbulent kinetic spectrum which may be measured by the  Parker Solar Probe (PSP) mission in the coming years, see Figure~\ref{fig:psp}.  This mission will be the first one approaching the Sun as close as 10~Solar Radii, i.e., $\sim 0.05$~AU. At such distances, the Alfv\'en speed is about $V_a\sim 300-1000$~km/s, \cite[e.g.,][]{Kasper2016SSR}, and the solar wind speed becomes sub-Alfv\'enic,  the magnetic field is very strong $B_0\sim 10^3$~nT  and plasma $\beta$ 
(the ratio between thermal and magnetic pressures) decreases down to $\beta\sim 0.03$ \cite{Bale2016SSR}. If the characteristic time of the solar wind  expansion  remains larger than the non-linear time of the eddies at kinetic scales, the turbulent spectrum at plasma kinetic scales will follow the general shape presented here. In physical units, the dissipation will happen at much smaller scales than at 1~AU, but still around the $\rho_e$ scale. See supplementary materials for more details.


\section{Nature of the kinetic plasma turbulence}


\begin{figure*}
\begin{center}
\includegraphics[width=20cm]{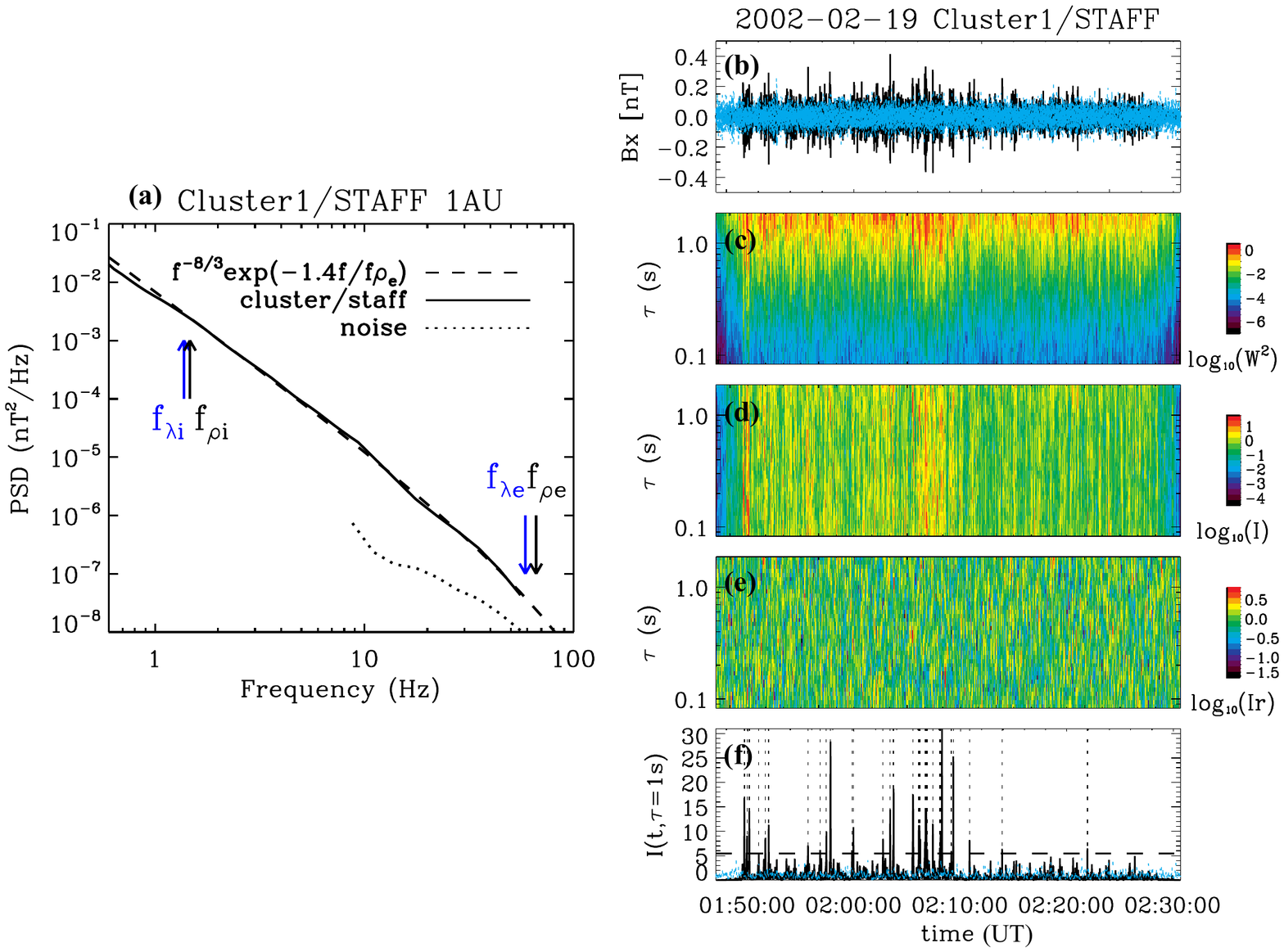} 
\caption{
(a) Spectrum of magnetic fluctuations at kinetic scales, as measured by Cluster1/STAFF at 1~AU on February 19, 2002. The ion (i.e., proton) and electron scales are indicated by arrows. 
The dashed line represents the fit with $\sim f^{-8/3}\exp{(-1.4f/f_{\rho e})}$. 
The dotted line is the instrumental noise. 
(b) The observed signal $B_x (t)$ in the GSE frame (black line), which forms the spectrum in (a), $B_{xr} (t)$ (blue line) has the same spectrum as $B_x (t)$, but random phases. 
(c) Wavelet scalogram $W^2 (t,\tau)$ of $B_x (t)$. 
(d) Normalised scalogram of $B_x (t)$, $I(t,\tau)$. 
(e) Normalised scalogram of $B_{xr} (t)$, $I_r (t,\tau)$. 
(f) Cuts of $I(t,\tau)$, in black, and $I_r (t,\tau)$, in blue, at timescale $\tau=1$~s. 
There is a clear difference between the observed and the random phase signal properties: 
time localised energetic events delocalised in timescales, visible in panel (d) as vertical lines, are destroyed by phase randomization, see panel (e). 
Thus, the observed spectrum in panel (a) is  primarily due to energetic events with coupled phases across scales.}
\label{fig:interm}
\end{center}
\end{figure*}

What kind of turbulence is `behind' this general spectrum? To answer this question, one needs to look at the fine structure of the fluctuations, responsible for the observed spectrum:  this requires high time-resolution measurements of kinetic scales.  Unfortunately, the Helios/SCM time-domain data are not available [F. Neubauer, private communication, 2016]. 
Thus, the time interval we have selected for our analysis is a typical one  at 1~AU from Cluster mission, and has been analysed previously at larger scales by \citet{bale05} and \citet{perrone16}.
Here, at kinetic scales, a spectrum similar to Figure~\ref{fig1} is observed, see Figure~\ref{fig:interm}(a). 
At frequencies below $\sim 10$~Hz, this spectrum was obtained using time series of the three components of magnetic fluctuations, as measured by Cluster/STAFF-SC in the normal mode (25 vectors per second).  One of these components, $B_x(t)$ in the GSE frame, is shown on Figure~\ref{fig:interm}(b) by a black line.

Figure~\ref{fig:interm}(c) shows the Morlet Wavelet scalogram $W^2(t,\tau)$ \cite{torrence98} of $B_x (t)$, i.e., the energy distribution of the signal in time $t$ and time-scales $\tau$ (or inversed frequencies $f$): it is non-homogeneously distributed  with localised stalactite-like events. Such an energy distribution is usually observed in the 
solar wind in absence of linear instabilities,  \cite[e.g.,][]{bale05,lion16,roberts16,perrone17}.

Figure~\ref{fig:interm}(d) gives the Local Intermittency Measure (\textit{LIM}), 
\begin{equation}\label{eq:lim}
I(t,\tau)=|W(t,\tau)|^2/\langle|W(t,\tau)^2|\rangle_t,
\end{equation}
of the observed signal \cite{farge92}. As one can see from the definition, \textit{LIM} 
allows to see deviations of the turbulent energy from its mean at each time-scale. 
In this $I(t,\tau)$-map, we observe  a high number of energetic events localised in time and delocalised in time-scales, forming vertical lines. 

\begin{figure}
\begin{center}
\includegraphics[width=7.5cm]{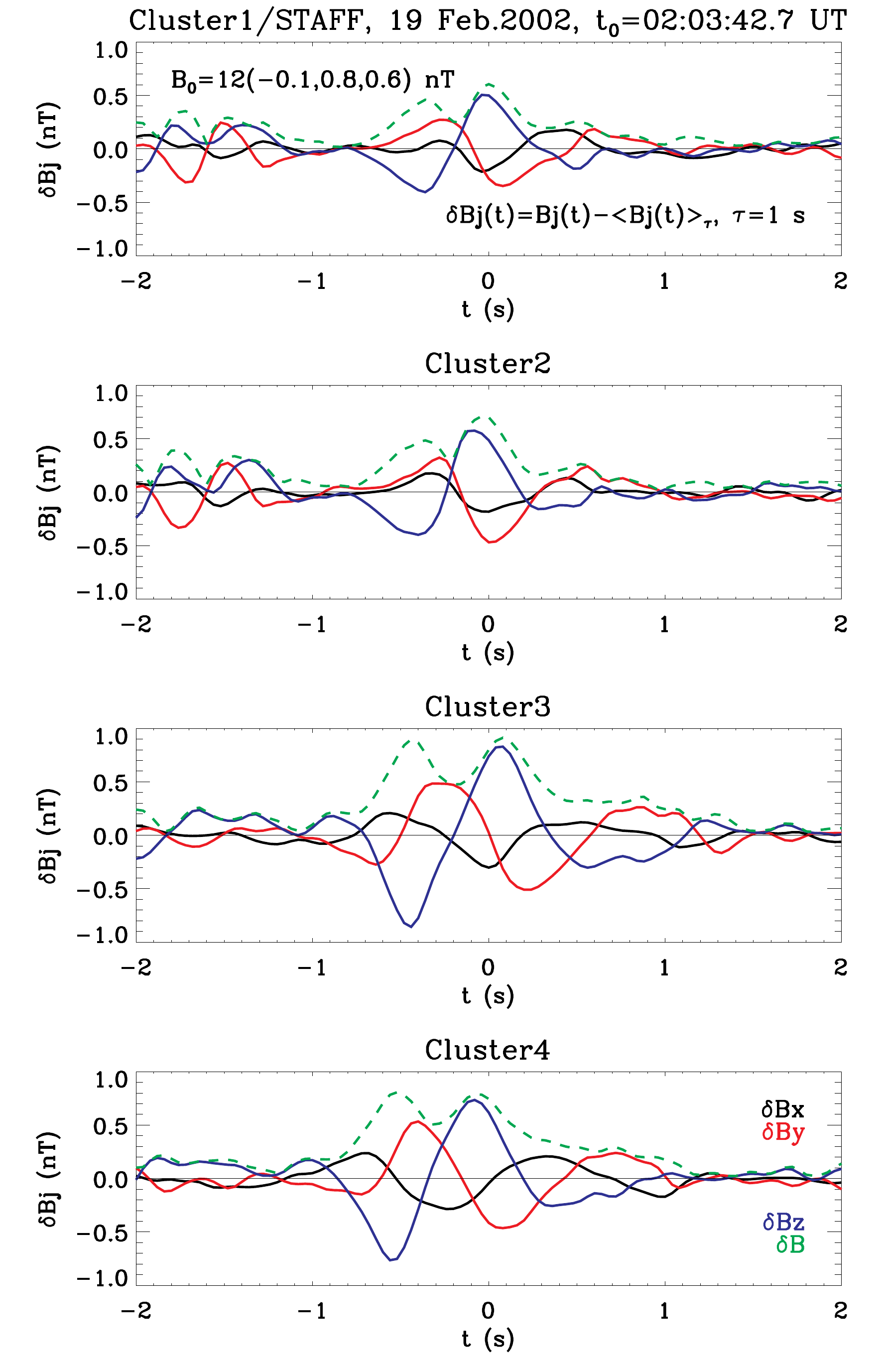}
\caption{ 
A vortex-like structure at sub-ion scales observed by the 4 Cluster satellites with inter-separation distances of about 200~km, during the time interval of Figure~\ref{fig:interm}. Magnetic field components are in the GSE frame. 
Such magnetic fluctuations correspond to current filaments localised in the centre of each structure with a cross section of the order of ion scales.  
 }
\label{fig:struc}
\end{center}
\end{figure}

What do these vertical lines in \textit{LIM} or stalactites in the Wavelet scalogram mean? 
We show below that they are signatures of coherent structures, for which wavelet coefficients have coupled phases across a wide range of time-scales. For this purpose, we perform the following numerical experiment, inspired by \citet{hada03}: we Fourier Transform  $B_x (t)$,  then we randomise phases keeping the amplitudes unchanged,  then we do the inverse Fourier Transform and get a random-phase signal $B_{xr} (t)$. 
This random phase signal is shown by the blue line on Figure~\ref{fig:interm}(b) and its intermittency measure  map, $I_r (t,\tau)$, in panel (e). One can see that the phase randomisation in the original signal destroys the extreme events in the time domain (panel b) and randomises the energy in the  $(t,\tau)$--plane (panel e): one does no longer observe the vertical lines in \textit{LIM}, corresponding to short duration energetic events, covering a wide range of time-scales. This leads us to conclude that all time localised and scale delocalised energetic events, observed in Figure~\ref{fig:interm}(d) have  phases which are coupled across a wide range of scales. Figure~\ref{fig:struc} shows magnetic fluctuations around such an event observed by the 4 satellites of Cluster (see the 4 panels): in the centre of the 4s-time interval we find coherent high amplitude fluctuations. The time delays between the satellites are consistent with a space localised cylindrical magnetic vortex at spatial scales of the order of the inter-satellite separations 
and which slightly  propagates ($\sim 0.4 V_a$) in the plasma frame quasi-perpendicularly to ${\bf B_0}$. The difference of amplitude of the fluctuations detected by different satellites indicates that the 4 satellites crossed the vortex  with slightly different trajectories, which confirms the space localisation of the structure.  
Similar vortices but at larger scales were previously observed by \citet{alexandrova06,perrone16,perrone17,roberts16}.


\begin{figure}
\begin{center}
\includegraphics[width=8cm]{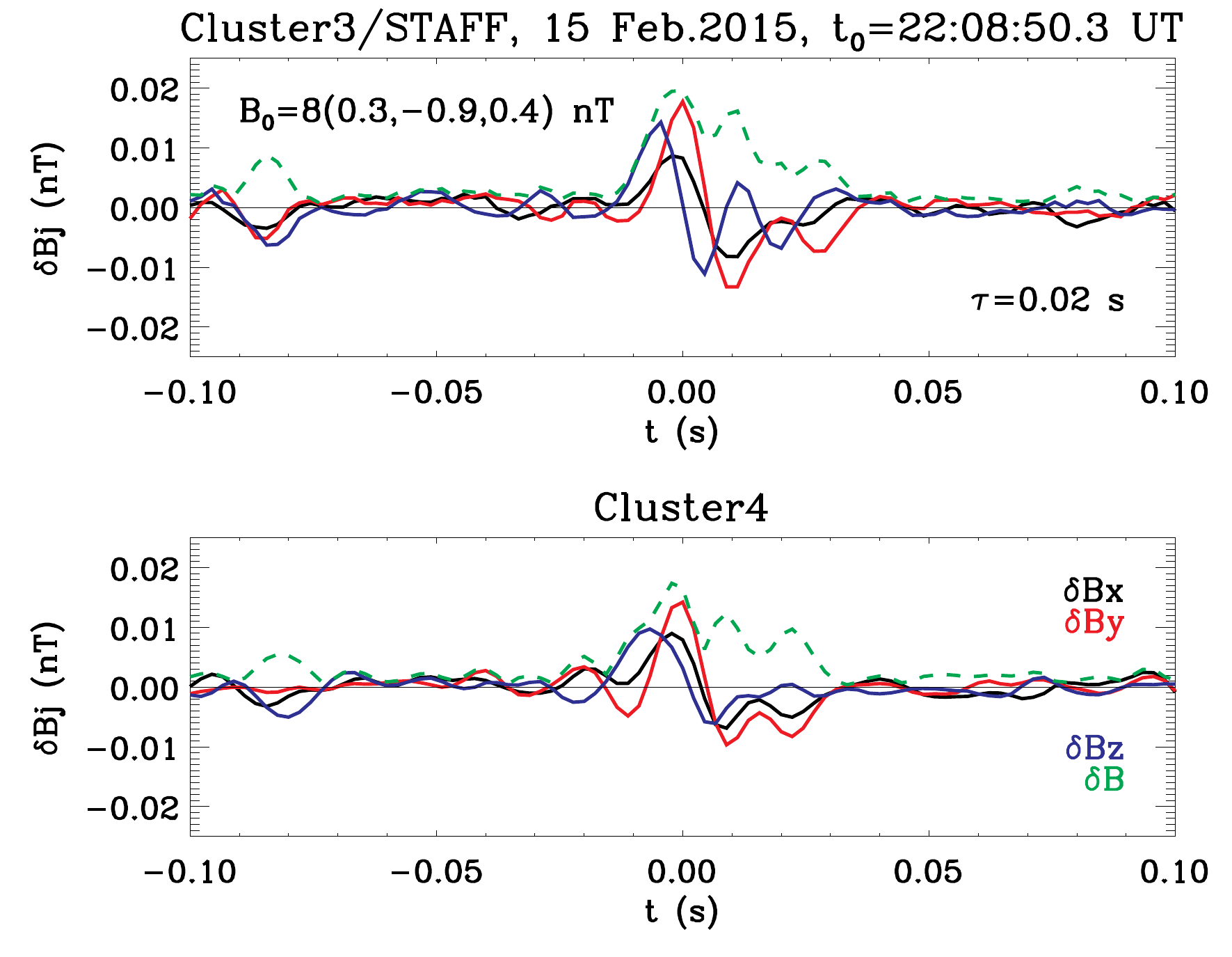} 
\caption{ 
Electron scale vortex-like structure crossed by 2 satellites of Cluster 7~km apart (Cluster Guest Investigator Operations); here the time interval is 20 times shorter than in  Figure~\ref{fig:struc}. 
Such magnetic fluctuations correspond to current filaments with a cross section of the order of several $\rho_e$.  
 }
\label{fig:struc-e}
\end{center}
\end{figure}

 During the typical time interval presented on Figures~\ref{fig:interm} and \ref{fig:struc}, the Cluster satellites were 200~km apart and Cluster/STAFF measures magnetic fluctuations with 0.04~s time resolution. Such measurements allow to observe ion and sub-ion scales but not electron scales. To resolve electron scales, we consider the data obtained during the Cluster Guest Investigator campaign of O. Alexandrova (2015-2016)~\footnote{See http://sci.esa.int/cluster/55616-guest-investigator-operations-2015-2016/}. 
The only available data in the free solar wind during this campaign is in the slow wind ($V\simeq 330$~km/s) on February 15, 2015. This time interval looks like any other typical solar wind turbulence, but here Cluster 3 (C3)  and Cluster 4 (C4) were only 7~km apart, and the time resolution is 0.0028s (i.e., 360 vectors per second), which allows  to resolve electron scales in space and in time simultaneously. 
We repeat the  above wavelet analysis on this time interval (not shown). 
It reveals results similar to Figure~\ref{fig:interm} in terms of non-homogeneous distribution of turbulence energy in the $(t,\tau)$--plane with energetic events localised in time and delocalised in time-scales, but here, up to the electron scales ($\tau \simeq 0.01$~s). 
The shape of the coherent structures at such small scales resembles magnetic vortices as well. 
An example of such an electron-scale magnetic vortex detected on 2 close satellites (C3 and C4) 
is shown  on Figure~\ref{fig:struc-e}: the duration of the crossing of such a vortex is about 0.05~s. 
The strongest gradient within this structure is localised within $\sim 0.01$~s, 
which corresponds to a spatial  scale of $\sim 3$~km, i.e., several electron Larmor radii $\rho_e$. 
 Note that this is the first time that such small-scale vortices are found in the solar wind. 
 They can be interpreted by the theory of electron-scale vortices in high-$\beta$ plasmas in the presence of an electron temperature anisotropy \cite{jovanovic15}. Similar structures have been found in 2D PIC numerical simulations \cite{haynes15} and in the Earth's plasma sheet \cite{sundberg15}; bigger magnetic vortices ($\sim 30 \rho_e$) have been recently detected by MMS in the Earth's magnetosheath \cite{huang17}. 
 

Thus,  for typical time intervals of solar wind turbulence, we observe an instance of \textit{strong turbulence}  at kinetic scales, i.e., with the presence of coherent structures with coupled phases across a wide range of scales,  namely, from ion to electron scales. 
 How general are these results at 1~AU? We have analysed in the same way a dozen of hours in total in the free solar wind between 2001 and 2006 under different plasma conditions and we have always found signatures of coherent structures at kinetic scales (see another example in supplementary materials C). Then, we have done a visual check of many random samples of STAFF measurements from 19 years of Cluster mission on the Cluster Quicklook (Fields \& Waves). Signatures of coherent structures, i.e., time localised and frequency delocalised energy enhancements, have been always present in the spectrograms while Cluster is in the free solar wind. 
 Thus, it seems that \textit{strong turbulence} is the typical situation at kinetic scales at 1~AU.
This implies that the dissipation is not expected to be homogeneous but is related to coherent structures. 
 
The topology of  the coherent structures is found here to be in the form of vortex-like filaments, like in neutral fluids and within the inertial range of the solar wind turbulence \cite{lion16,roberts16,perrone16,perrone17}. Within the kinetic range, previously, only small scale current sheets have been found \cite{greco16,perri12a}. 
An interesting task will be to estimate the filling factor of vortices and current sheets at such small scales.  A larger statistical study of kinetic scale vortices  in the solar wind and a possible relation with coherent structures observed within the inertial range  (embedding or just continuation of the same structures) will be the subject of future studies.

Can we say  that what we observe at 1~AU is typical for the Heliosphere? 
 The first results of PSP at $\sim 0.17$~AU \cite{Bale2019Nat} show that during time intervals which measures fluctuations with $\mathbf{k} {\perp} \mathbf{B_0}$ (for a non-radial-field wind), the same signatures of coherent structures  covering inertial and kinetic ranges are present, see Figure~3 in \cite{Bale2019Nat}. 
 What is the topology of these structures closer to the Sun? How do they evolve with radial distance? 
 What is their life-time? and how can it be taken into account by  turbulence models?
We will address these questions in future studies using Parker Solar Probe and Solar Orbiter measurements.

\section{Conclusions and discussions }

Despite important differences between collisionless plasmas and neutral fluids,  we find that 
they develop turbulent states which present several qualitative similarities.

It was already known that at MHD scales, magnetic spectra follow the same shape $\sim k^{-5/3}$ for different radial distances from the Sun \cite{bruno13}. Here, we show that at smaller scales the spectrum keeps also its shape 
independently of the radial distance (from 0.3 to 1~AU), with an exponential fall-off reminiscent of the dissipation range of neutral fluid turbulence $\sim f^{-8/3} \exp{(-f/f_{d})}$.    
We show as well that the equivalent of the Kolmogorov scale $\ell_d$, where the dissipation of the electromagnetic cascade is expected to  take place, is controlled by the electron Larmor radius  $\rho_e$ for different radial distances.

This is not a trivial result, since the electron Larmor radius is not the only characteristic length at such small scales. Closer to the Sun, the electron inertial length $\lambda_e$ becomes larger than the Larmor radius $\rho_e$, but, as observed here, it is still with $\rho_e$ and not with $\lambda_e$  that the ``dissipation'' scale correlates.  

A significant difference occurs here with neutral fluids turbulence. In neutral fluids, the dissipation scale $\ell_d$ depends on the energy injection and is  much larger than the mean free path, so that the dissipation range is described within the fluid approximation.  As we show here, in the solar wind between 0.3 and 1~AU,  on the contrary, $\ell_d$  is defined by  $\rho_e$ independently of the rate of energy injection.  In the vicinity of $\rho_e$ the protons are completely kinetic and electrons start to be kinetic.

The nature of the turbulent fluctuations which form the observed spectrum, presents also some similarities
 with the neutral fluid turbulence situation, as we observe at 1~AU. 
Despite small amplitudes at kinetic scales with respect to the mean field, we  find 
many coherent structures, which cover a wide range of scales but are localised in space.  These coherent events look like magnetic vortex filaments with cross section of the order of  several $\rho_e$, and may play a crucial role in the dissipation of the space plasma turbulent cascade. 

The observed general features and similarities with the usual fluid turbulence seem to indicate that similar physical processes are likely to be universal for the whole Heliosphere and may also turn out to be universal for other astrophysical turbulent plasma environments such as the interstellar medium or magnetospheres of other stars.

\section{Supplementary materials}

\subsection{Helios-1 spectra}

Measurements of  magnetic fluctuations at kinetic scales are challenging: here, the level of background turbulence is low therefore only very sensitive instruments can capture the corresponding fluctuations. The most sensitive instrument devoted to kinetic scales at the moment is the Cluster/STAFF, doing measurements at 1~AU.  
Approaching the Sun, the turbulence level increases and therefore even with less sensitive instruments, one may expect to observe kinetic spectrum up to electron scales. 

The Helios/SCM instrument \cite{Neubauer1977-zg}  provides magnetic spectra for two of three components, $(B_y, B_z)$ and rarely $(B_x, B_z)$, in the Spacecraft Solar ecliptic reference frame, which is equivalent to the Geocentric Solar Ecliptic (GSE) frame. 
For the present study we use the spectra of $B_y$ only. Indeed, the pre-flight noise level for the $B_y$ spectra matches well with the post-flight noise level, that is not the case for $B_z$ \cite{Neubauer1977-jgr}. The SCM spectra are integrated over 8 seconds, for 8 logarithmically spaced frequency bands, with central frequencies going from 7~Hz up to 1.5~kHz. No time domain measurements are available. Therefore, it is impossible to restore the errors on the integrated spectra over 8 seconds.


\begin{figure}
\begin{center}
\includegraphics[width=7.5cm]{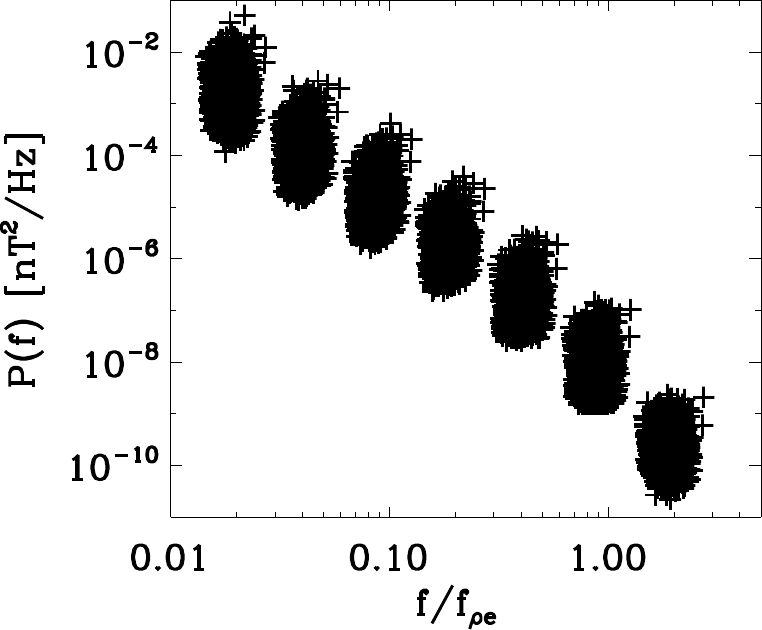}
\caption{The same 3344 spectra (0.3~AU) as in Figure~\ref{fig1}A(top), but  cleaned from the noise and normalised by the Doppler shifted $\rho_e$, $f_{\rho e}$.}
\label{fig:helios_ana}
\end{center}
\end{figure}
 Below, we explain how Figure~\ref{fig1}A(bottom) was obtained.  
As we discuss in the main body of the paper, at  0.3 AU, in the fast wind,  there are 3344 spectra with signal-to-noise ratio (SNR) larger than 3 up to 316~Hz, and with plasma measurements in the close vicinity of the spectra (i.e., mean field at most within 16 seconds around the measured SCM spectrum,  electron temperature within about 30 minutes and when not available, it is taken within the longer time interval from  the same wind type), see Figure~\ref{fig1}A(top). 
 The subset of 39 most intense spectra with SNR$>3$ up to $681$~Hz and  
with  simultaneous measurements of $\mathbf{B_0}$
(see green crosses in Figure~\ref{fig1}A(top))  was used to determine two out of three free parameters  of the model function $A f^{-\alpha}\exp{(-f/f_{d})}$, namely $\alpha=-8/3$ and $f_d=f_{\rho e}/1.8$. Now, let us verify if the established model $A f^{-8/3}\exp{(-1.8f/f_{\rho e})}$ describes well the rest of the data. Figure~\ref{fig:helios_ana} shows the same 3344 spectra as in Figure~\ref{fig1}A(top), 
but here the spectra are cleaned from the noise, $P(f)=PSD(B_y)-PSD_{noise}$, and frequencies are normalised by the Doppler shifted electron Larmor radius, $f_{\rho e}$. 
Figure~\ref{fig1}A(bottom) shows 2D histogram calculated with the same spectra as in 
Figure~\ref{fig:helios_ana}
but which all passe by one point  $(f_0,P_0)=(0.051,10^{-4})$, i.e., collapsed in amplitude around $f/f_{\rho e}=0.051$  (the results do not change if we choose another way to collapse the spectra). The dashed line indicates  the function $A f^{-8/3}\exp{(-1.8 f/f_{\rho e})}$, which passes through the data without any particular adjustment. Note, that the  dispersion of the data points at lowest and highest frequency ends can be due to the non-simultaneous $T_e$ measurements. Moreover, the 
lowest frequency can be affected  as well by the presence of the ion characteristic scales; and the highest frequencies -- by the SCM noise.

Further from the Sun, the intensity of magnetic fluctuations decreases following the mean field $B_0$, \cite[e.g.,][]{beinroth81,bourouaine12},  
 but the kinetic scales increase (i.e., characteristic frequencies decrease) so we can resolve them with less frequency bands: At 0.6 AU, we have $\sim 3000$ 
 spectra up to 147~Hz and 21 454 spectra up to 68~Hz; 
 at 0.9 AU, there are 24 spectra  with SNR$>3$ up to 147~Hz and 10570 spectra up to 68~Hz.  
  Their shape is still similar to what is observed at 0.3~AU.

\subsection{Extrapolation of turbulent spectra closer to the Sun}

Figure~\ref{fig:psp} of the paper shows the complete turbulent spectrum covering the energy containing scales ($\sim f^{-1}$ spectral range), the inertial range of MHD scales
($\sim f^{-5/3}$ range) and the kinetic scales, as observed at 0.3 and 0.9~AU by Helios. 
The most intense spectra (in green and red) are the predictions for 
PSP at 0.1 and 0.05~AU, respectively. 
For this, we assume that the turbulence level will increase together with the mean field, keeping $\delta B/B_0\sim const$, as observed in the solar wind,  \cite[e.g.,][]{beinroth81}.
The onset of the Kolmogorov inertial range  is assumed to start at the frequency $f_b$ (black dots), which varies with $R$ as $f_b= f_0 (R_0/R)^{1.52}$ as is the case for the observed turbulence between 0.3 and 5 AU \cite{bruno13}. 
In the inner heliosphere, where $\beta<1$, the end of the Kolmogorov scaling is expected to happen at the proton inertial length $\lambda_p$,  \cite[e.g.,][]{bourouaine12}.  The exponential roll-off at the end of the electromagnetic cascade is defined by the local $\rho_e$, as we confirm in this study. 
To determine the Doppler shifted frequencies where  $\lambda_p$ and $\rho_e$ will appear in the spectra ($f_{\lambda p}=V/2\pi \lambda_p$   and $f_{\rho e}=V/2\pi \rho_e$), we use plasma parameters (density $n$, electron temperature $T_e$, magnetic field $B_0$ and solar wind speed $V$) extrapolated from the in-situ Helios measurements (from 0.3 to 0.9 AU) [Maksimovic et al., in preparation]. 
More precisely these latter extrapolations have been performed by connecting the gradient of the Helios density measurements to the one measured remotely from coronal white light eclipse observations. 
In addition, the bulk speed profiles have been obtained by imposing the conservation of mass flux all the way down to one solar radius. 
The plasma parameters used for the predicted spectra as well as for the time intervals of the Helios measurements are summarized in Table~1.

\begin{table}
\centering
\caption{Mean plasma parameters at 4 radial distances from the Sun, corresponding to the spectra in Figure~\ref{fig:psp}.}
    \begin{tabular}{ c | c c c c c c c c c }
    \hline
     & 0.9 AU& 0.3 AU & 0.1 AU& 0.05 AU  \\ \hline
  
    $B_0$ (nT) & $7\pm2$& $41\pm 3$ & 280 & 990  \\ 
      $V$ (km/s)  & $705\pm35$ & $650\pm 40$ & 510 & 410 \\ 
   
    $N$ (cm$^{-3}$) & $4\pm1$ & $31\pm 4$ & 350 &1700 \\ 
    
   $T_p$ (eV) &$21\pm 5$ &$50\pm9$ & 120 & 230 \\ 
   $T_e$ (eV) &$9 \pm 2$ &$15\pm 2$ & 19 & 25 \\ 
   $T_{p \perp}$ (eV) &$24\pm 5$ &$65\pm10$ & - & - \\ 
   $T_{e \perp}$ (eV) &$7\pm1$ &$12\pm1$ & - & -  \\ 
   $\beta_p$ & $0.8\pm 0.2$ & $0.5\pm 0.1$ & 0.2 & 0.15 \\ 
   $\beta_e$  &$0.2\pm0.1$ &$0.10\pm0.02$ & 0.04 & 0.02 \\ 
   $\lambda_p$ (km) &$108 \pm 14$ &$39\pm 3$ & 12 & 6  \\ 
   $\rho_p$ (km) &$101 \pm 31$ &$28\pm 3$& 6 & 2 \\ 
   $\lambda_e$ (km)  &$2.5 \pm 0.3$ &$0.9 \pm 0.1$ & 0.3 & 0.1  \\ 
   $ \rho_e$ (km) &$1.3\pm 0.4$ &$0.3 \pm 0.02$ & 0.05 & 0.02  \\ 
   $f_{cp}$ (Hz) &$0.10 \pm0.03 $&$0.6 \pm 0.05$ & 4 & 15  \\ 
   $f_{\lambda p}$ (Hz)  &$1.0 \pm 0.1$& $2.6 \pm 0.3$ & 7 & 12 \\ 
   $f_{\rho p}$ (Hz) &$1.1 \pm 0.3$ &$3.6 \pm 0.5$ & 14 & 30  \\ 
   $f_{\lambda e}$ (Hz)  & $44\pm 6$&$110 \pm 10$ & 300 & 500  \\ 
   $f_{\rho e}$ (Hz) &$90 \pm 30$ &$360\pm 40$ & 1530 & 3800 \\ 
   $f_{ce}$ (Hz) &$200 \pm 60 $&$1150\pm 80$ & 7800 & 28000  \\ \hline
   \end{tabular}
\end{table}

It is possible that we overestimate the predicted spectrum at 0.05~AU. 
In fact, in the Heliosphere, $\delta B/B_0$  is of the order of unity at the largest scales  of the cascade, around $f_b$  \cite{matteini18}. 
But close to the Alfvén point (where $V=V_a$) in the vicinity of 0.05~AU, this ratio is expected to be of the order of 0.3 \cite{verdini07}. 
Therefore, the  spectrum can be about one order of magnitude lower than presented here. 
In the coming years, PSP measurements close to the Sun will show how the empirical picture of the turbulence given in this article may change.

\subsection{Signatures of coherent structures within the kinetic range: a fast wind example}

\begin{figure}
\begin{center}
\includegraphics[width=7cm]{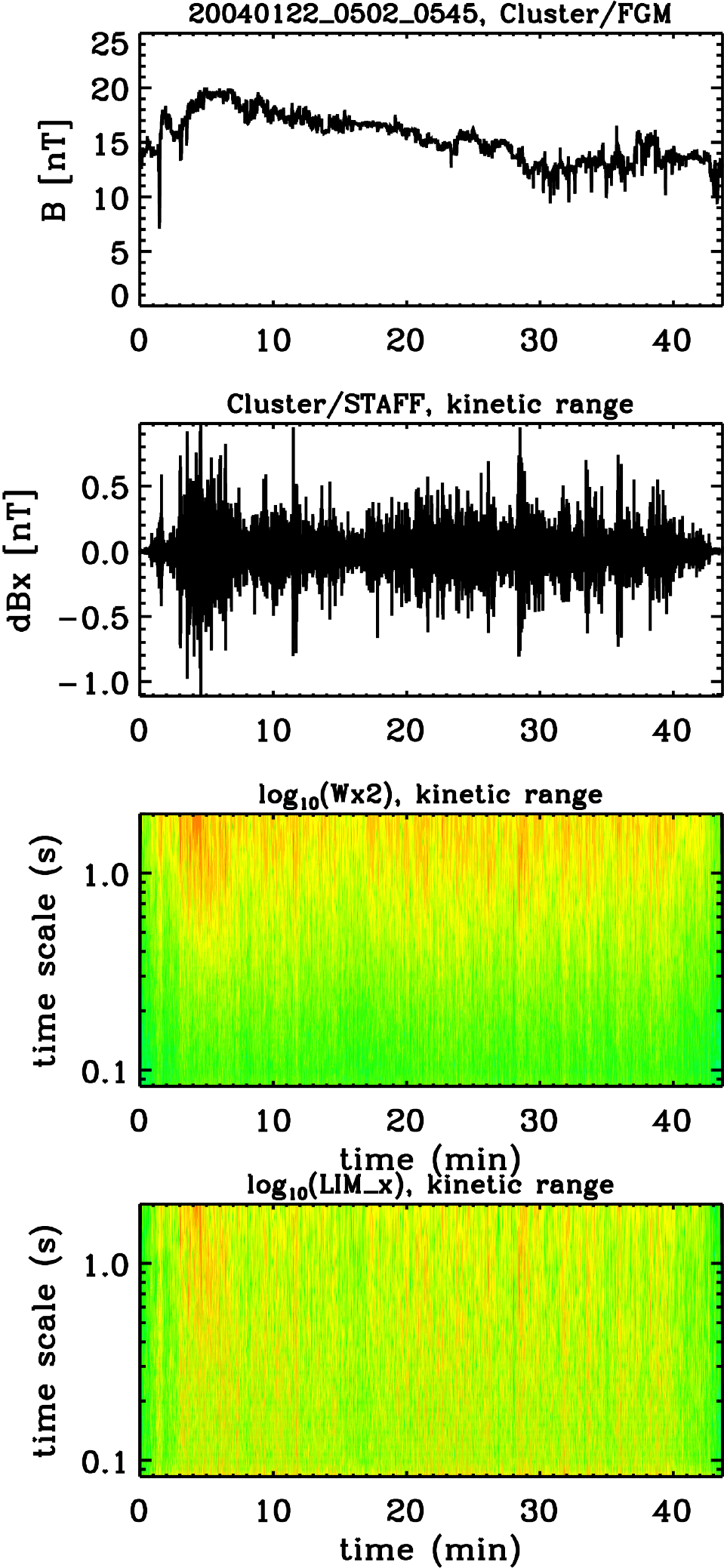}
\caption{Magnetic field in the fast solar wind on January 22, 2004. From up to the bottom: norm of the large scale magnetic field as measured by Cluster/FGM; 
$B_x$ component of magnetic fluctuations in the GSE reference frame, within the kinetic range of scales (Cluster/STAFF-SC); 
Wavelet scalogram of $B_x$; 
\textit{LIM} of $B_x$. 
Vertical lines in \textit{LIM} are signatures of coherent structures at kinetic scales.}
\label{fig:cluster_2004}
\end{center}
\end{figure}

The results presented in section 3 of the paper  seems to be typical for solar wind turbulence  at 1 AU in the absence of signatures of linear instabilities such as Alfv\'en Ion Cyclotron (AIC) waves at ion scales and whistler waves at electron scales. These linear waves represent only few percents of solar wind data at 1~AU \cite{Jian2014ApJ,lacombe14}.  In the rest of the data, the typical kinetic spectrum is observed (see section 2 of the paper).  For such time intervals,  we usually observe 
signatures of intermittent coherent structures, i.e., the wavelet decomposition shows localised events in time and delocalised in scales.  In section 3 of the paper, we have shown examples from the slow wind streams.  
Figure~\ref{fig:cluster_2004} gives an example of the fast wind ($V\sim 670$~km/s). Kinetic turbulent  spectrum for this time interval follows the general shape and can be found in Figure~2 of \cite{alexandrova12}. Two bottom panels of Figure~\ref{fig:cluster_2004} give  the wavelet scalogram and the Local Intermittency Measure (\textit{LIM}) of magnetic field fluctuations in the kinetic range of scales. 
One observes here the same  signatures of coherent structures  as in the slow wind (Figure~\ref{fig:interm}(c,d)): stalactites in $W^2(\tau,t)$ and vertical lines in the \textit{LIM}.  This is a typical picture for 1 AU. It will be interesting to verify these results with the new measurements of Parker Solar Probe and  Solar Orbiter data closer to the Sun.

\paragraph*{Acknowledgement}
 OA, VJ and MM are supported by the French Centre National d’Etude Spatiales (CNES). 
 OA thanks C. Lacombe and L. Matteini for discussions.
\paragraph*{Data} 
The Helios-1 data are available on the Helios data archive (http://helios-data.ssl.berkeley.edu/). 
The Cluster data are available on the Cluster Science Archive (https://csa.esac.esa.int/csa-web/). 
\paragraph*{Software} 
The Wavelet software was provided by C. Torrence and G. Compo and is available at http://paos.colorado.edu/research/wavelets/.
 \paragraph*{Correspondence} Correspondence 
should be addressed to O. Alexandrova~(email: olga.alexandrova@obspm.fr).

\bibliography{nature_turbu_2019}

\end{document}